\documentclass[
aps,
prl,
longbibliography,
superscriptaddress,
floatfix,
reprint
]{revtex4-2}

\usepackage{mathrsfs}
\usepackage{amsmath}
\usepackage{amsfonts}
\usepackage{amssymb}
\usepackage{amsthm}
\usepackage{graphicx}
\usepackage{color}
\usepackage[pdfpagelayout=OnePageLeft]{hyperref}
\usepackage{bm}
\usepackage[caption=false]{subfig}
\usepackage{verbatim}
\usepackage[
per-mode=symbol,
]{siunitx}
\usepackage{booktabs}
\usepackage{xspace}

\DeclareSIUnit\dBm{dBm}
\DeclareSIUnit\dB{dB}
\DeclareSIUnit\inch{in}

	\if@noloadpdfcomment
	\newcommand{\todo}[1]{%
		\marginpar{#1}%
		\if@draftmode
		\relax
		\else
		\ClassWarning{main}{Warning! There are \string\todo\space notes in the final document.}
		\fi}
	\else
	\RequirePackage{pdfcomment}
	\newcommand{\todo}[1]{%
		\pdfcomment[open=true]{#1}%
		\if@draftmode
		\relax
		\else
		\ClassWarning{main}{Warning! There are \string\todo\space notes in the final document.}
		\fi}
	\fi

\begin{document}
\title{Tunable Cooperativity in Coupled Spin--Cavity Systems}

\author{Lukas~Liensberger}
\email[]{Lukas.Liensberger@wmi.badw.de}
\affiliation{Walther-Mei{\ss}ner-Institut, Bayerische Akademie der Wissenschaften, 85748 Garching, Germany}
\affiliation{Physik-Department, Technische Universit\"{a}t M\"{u}nchen, 85748 Garching, Germany}

\author{Franz~X.~Haslbeck}
\affiliation{Lehrstuhl f\"{u}r Topologie korrelierter Systeme, Physik-Department, Technische Universit\"{a}t M\"{u}nchen, 85748 Garching, Germany}

\author{Andreas~Bauer}
\affiliation{Lehrstuhl f\"{u}r Topologie korrelierter Systeme, Physik-Department, Technische Universit\"{a}t M\"{u}nchen, 85748 Garching, Germany}

\author{Helmuth~Berger}
\affiliation{Institut de Physique de la Mati\`{e}re Complexe, \'{E}cole Polytechnique F\'{e}d\'{e}rale de Lausanne, 1015 Lausanne, Switzerland}

\author{Rudolf~Gross}
\affiliation{Walther-Mei{\ss}ner-Institut, Bayerische Akademie der Wissenschaften, 85748 Garching, Germany}
\affiliation{Physik-Department, Technische Universit\"{a}t M\"{u}nchen, 85748 Garching, Germany}
\affiliation{Munich Center for Quantum Science and Technology (MCQST), 80799 Munich, Germany}

\author{Hans~Huebl}
\affiliation{Walther-Mei{\ss}ner-Institut, Bayerische Akademie der Wissenschaften, 85748 Garching, Germany}
\affiliation{Physik-Department, Technische Universit\"{a}t M\"{u}nchen, 85748 Garching, Germany}
\affiliation{Munich Center for Quantum Science and Technology (MCQST), 80799 Munich, Germany}

\author{Christian~Pfleiderer}
\affiliation{Lehrstuhl f\"{u}r Topologie korrelierter Systeme, Physik-Department, Technische Universit\"{a}t M\"{u}nchen, 85748 Garching, Germany}
\affiliation{Munich Center for Quantum Science and Technology (MCQST), 80799 Munich, Germany}

\author{Mathias~Weiler}
\email[]{weiler@physik.uni-kl.de}
\affiliation{Walther-Mei{\ss}ner-Institut, Bayerische Akademie der Wissenschaften, 85748 Garching, Germany}
\affiliation{Physik-Department, Technische Universit\"{a}t M\"{u}nchen, 85748 Garching, Germany}
\affiliation{Fachbereich Physik and Landesforschungszentrum OPTIMAS, Technische Universit\"{a}t Kaiserslautern, 67663 Kaiserslautern, Germany}

\date{\today}

\begin{abstract}
We experimentally study the tunability of the cooperativity in coupled spin--cavity systems by changing the magnetic state of the spin system via an external control parameter. As model system, we use the skyrmion host material Cu$_2$OSeO$_3$ coupled to a microwave cavity resonator. In the different magnetic phases we measure a dispersive coupling between the resonator and the magnon modes and model our results by using the input--output formalism. Our results show a strong tunability of the normalized coupling rate by magnetic field, allowing us to change the magnon--photon cooperativity from 1 to 60 at the phase boundaries of the skyrmion lattice state.
\end{abstract}

\maketitle

Strongly coupled magnon--photon systems become particularly interesting, when the magnitude of the effective coupling rate $g_\mathrm{eff}$ between the magnons and photons in a microwave resonator exceeds their respective loss rates $\kappa_\mathrm{mag}$ and $\kappa_\mathrm{res}$. In this limit, coherent exchange of the quantized excitations is established~\cite{Lachance-Quirion2019,FriskKockum2019} and for instance has been extensively studied for magnon--photon coupling in ferromagnets and ferrimagnets~\cite{Huebl2013,Zhang2014,Bai2015,Liu2016,Harder2018}. A measure of the coherent exchange of excitations in spin-cavity systems is the cooperativity. The possibility to tune the cooperativity by tiny changes of an external control parameter is interesting by itself and a promising tool for applications. A possible route towards large tunability is offered by inducing changes in the spin system. To experimentally demonstrate this concept, we studied the skyrmion host material Cu$_2$OSeO$_3$, featuring several magnetic phases in a small magnetic field and temperature window. 

Skyrmions are topologically stabilized spin solitons~\cite{Muhlbauer2009,Jonietz2010,Adams2012,Schulz2012,Seki2012,Yu2012,Fert2013,Nagaosa2013,Garst2017,Kanazawa2017,Bauer2018} exhibiting a rich spectrum of collective excitations of the spin system known as magnons. Theses excitations have been extensively studied in the limit of weak coupling to microwave circuits~\cite{Onose2012, Okamura2013, Schwarze2015, Okamura2015, Weiler2017, Stasinopoulos2017, Aqeel2021}. For potential applications such as realizing a novel magnetic state storage~\cite{Li2020}, it is necessary to couple the magnons to other quantized excitations such as photons~\cite{Tabuchi2015} or (quasi-)particles. While the coupling of the higher-order helimagnon modes~\cite{Weiler2017} to an X-band (\SI{9.8}{\giga\hertz}) photonic resonator has been demonstrated already in Cu$_2$OSeO$_3$~\cite{Abdurakhimov2019}, the coupling in the skyrmion lattice phase and the potential for tunable cooperativity associated with it remained an open issue so far.

In this study, we report the coupling of uniform-mode excitations of the insulating chiral magnet Cu$_2$OSeO$_3$ to microwave photons in a three-dimensional microwave cavity with a resonance frequency of \SI{683.1}{\mega\hertz}. In contrast to earlier experiments~\cite{Abdurakhimov2019}, the magnon--photon coupling was addressed in all magnetic phases, including the skyrmion lattice phase. We find a large coupling rate of the breathing-mode skyrmion excitation to the cavity photons resulting in a high cooperativity $C=g_\mathrm{eff}^2/(\kappa_\mathrm{mag}\,\kappa_\mathrm{res})>50$. Most importantly, the cooperativity can be tuned to values as low as 1 by changing the magnetic field by only $\sim \SI{10}{\milli\tesla}$.

\begin{figure}[h]
	\begin{center}
		\includegraphics[]{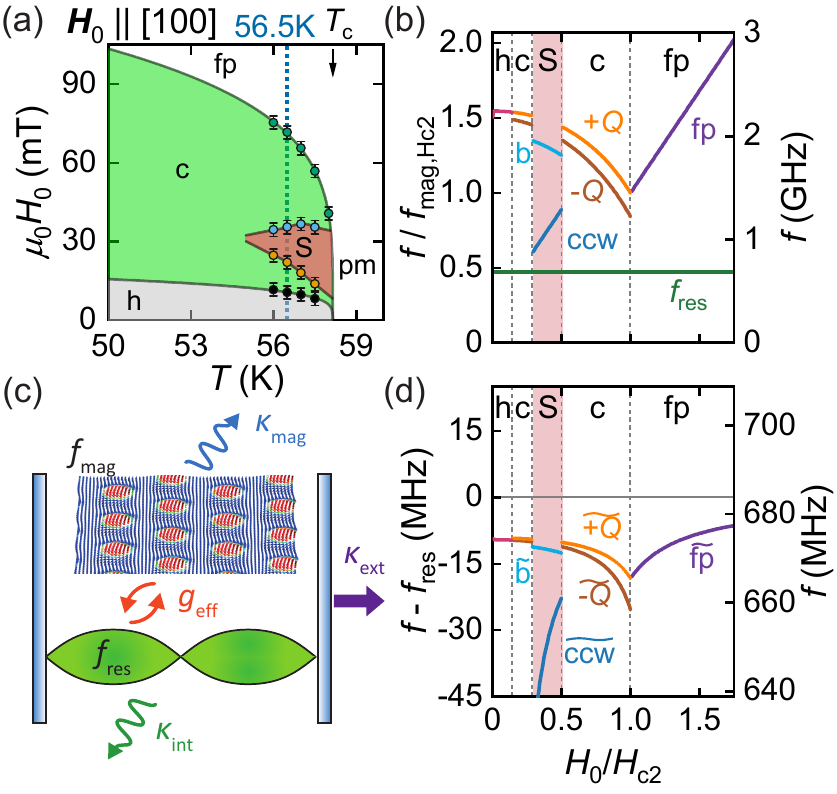}
		\caption{Microwave dynamics in Cu$_2$OSeO$_3$. (a)~Magnetic phase diagram extracted from broadband magnetic resonance measurements. The following magnetic phases are distinguished: field-polarized~(\textbf{fp}), conical~(\textbf{c}), helical~(\textbf{h}) phase, skyrmion lattice phase~(\textbf{S}), paramagnetic~(\textbf{pm}) phase. (b)~Schematic magnetic field dependence of the uncoupled, uniform resonance frequencies of the magnon modes $f_\mathrm{mag}$ in a chiral magnet. The resonance frequency of the resonator $f_\mathrm{res}$ is lower than all magnon frequencies. Normalized and absolute frequencies are shown on the left and right ordinate, respectively. (c)~Illustration of the input-output formalism and the parameters of Eq.~\eqref{eq:CSO_InputOutput}. (d)~Frequency shift of the resonator mode in a coupled magnon--photon system as calculated using the input--output formalism with constant coupling rate $g_\mathrm{eff}/(2\pi)=\SI{120}{\mega\hertz}$ in all magnetic phases. The hybridized magnon--cavity modes are denoted by tildes. }
		\label{fig:CSO_general}
	\end{center}
\end{figure}

Cu$_2$OSeO$_3$ possesses a non-centrosymmetric cubic crystal structure with space group $P$2$_1$3~\cite{Seki2012, Garst2017} and exhibits a rich magnetic phase diagram~\cite{Bauer2018} schematically depicted in Fig.~\ref{fig:CSO_general}(a). Below the critical temperature $T_\mathrm{c}\approx\SI{58}{\kelvin}$, the long-range ordered helical, conical, and skyrmion phases form. At large magnetic fields, the system is in the field-polarized state. 
Above the transition temperature $T_\mathrm{c}$ the system is paramagnetic. The phase diagram in Fig.~\ref{fig:CSO_general}(a) for $\bm{H}_0 \parallel [100]$ is representative for different crystallographic directions due to the rather weak magnetocrystalline anisotropy. In this study, the investigation of magnon--photon coupling was performed at $T=\SI{56.5}{\kelvin}$ as indicated by the dashed blue line in Fig.~\ref{fig:CSO_general}(a).

The different magnetic phases also differ in their magnetization dynamics~\cite{Onose2012,Schwarze2015,Garst2017}. In Fig.~\ref{fig:CSO_general}(b), the typical spectrum is shown for the uniform resonance frequencies of the fundamental modes of an insulating chiral magnet as a function of the external magnetic field. Starting at large fields, in the field-polarized phase~(\textbf{fp}) only one resonance mode is observed. Here, all the magnetic moments precess in-phase also referred to as the ferrimagnetic resonance. By decreasing the magnetic field, the system enters the conical phase~(\textbf{c}) where, depending on the details of the demagnetization, the $\pm\mathrm{Q}$ modes are observed~\cite{Schwarze2015}. In the skyrmion phase~(\textbf{S}), the breathing mode (b), the counterclockwise (ccw) and the clockwise mode can couple to a uniform driving field~\cite{Mochizuki2012}. The clockwise mode is omitted due to its comparably small spectral weight. For details on the characteristics of these excitation modes in chiral magnets, we refer to Ref.~\cite{Schwarze2015}.

To model the dynamics of the coupled chiral magnet--cavity system we employ the input--output formalism~\cite{Walls2008, Clerk2010, Schuster2010, Abe2011, Huebl2013}. A schematic of the physical mechanisms underlying the coupling and its description is shown in Fig.~\ref{fig:CSO_general}(c). The photons in a microwave resonator with frequency~$f_\mathrm{res}$ can interact with the magnons of the chiral magnet with frequency $f_\mathrm{mag}$ (via the dipolar interaction) with an effective coupling strength~$g_\mathrm{eff}$. The cavity photons and the magnons in Cu$_2$OSeO$_3$ possess certain lifetimes and thus decay with loss rates $\kappa_\mathrm{res}$ and $\kappa_\mathrm{mag}$, respectively. Note that the total loss rate $\kappa_\mathrm{res}$ is given by the sum of the internal loss rate $\kappa_\mathrm{int}$ and the external loss rate $\kappa_\mathrm{ext}$ of the cavity. The latter quantifies the coupling strength of the cavity to the feedline. For the measured $S_{11}$ parameter of the vector network analyzer (VNA) the input--output formalism yields~\cite{Walls2008, Clerk2010, Abe2011, Abdurakhimov2019}:
\begin{equation}
S_{11}(f) = -1 + \frac{\kappa_\mathrm{ext}}{\imath\cdot 2\mathrm{\pi} (f_\mathrm{res}-f) + \kappa_\mathrm{res} + \frac{g_\mathrm{eff}^2}{\imath \cdot 2\pi(f_\mathrm{mag}-f)+\kappa_\mathrm{mag}}},
\label{eq:CSO_InputOutput}
\end{equation}
with $\kappa_\mathrm{res}=\kappa_\mathrm{int}+\kappa_\mathrm{ext}$. The resonance frequency of the cavity, $f_\mathrm{res}=\SI{683.8}{\mega\hertz}$ at $\SI{56.5}{\kelvin}$, is lower than all resonance frequencies of the magnon modes, c.f. conventional broadband magnetic resonance spectroscopy in the Supplemental Material~\cite{SupInf}. This significant detuning of the two systems restricts our analysis to the dispersive limit~\cite{Abdurakhimov2019}. Using the resonance frequencies $f_\mathrm{mag}$ and $f_\mathrm{res}$ shown in Fig.~\ref{fig:CSO_general}(b) and $g_\mathrm{eff}/(2\pi)=\SI{120}{\mega\hertz}$ (compatible to the values we find below), we calculate the modified resonance frequencies of the coupled magnon--cavity system close to \SI{680}{\mega\hertz} by minimizing~$|S_{11}(f)|$ given by Eq.~\eqref{eq:CSO_InputOutput} for each magnetic field~$H_0$.
The result is shown in Fig.~\ref{fig:CSO_general}(d). In all magnetic phases of Cu$_2$OSeO$_3$, the frequency of the hybridized magnon-photon mode is reduced to values well below the resonance frequency $f_\mathrm{res}$ of the unperturbed cavity due to the formation of normal modes. 

\begin{figure}[t]
	\begin{center}
		\includegraphics[]{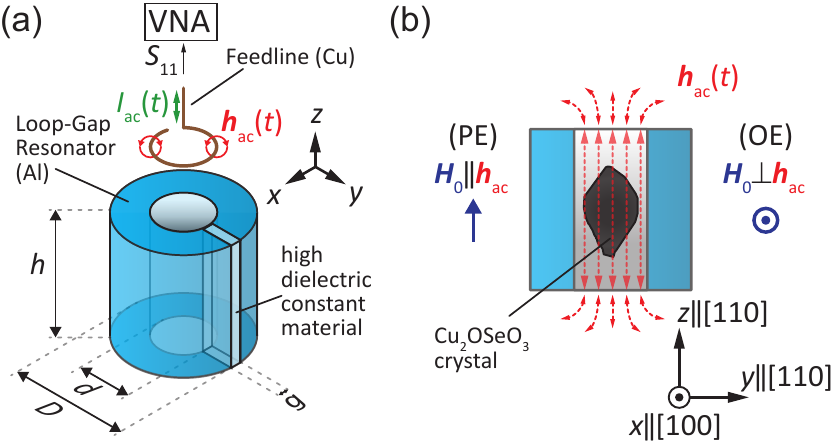}
		\caption{Experimental setup. (a)~Schematic of the loop--gap resonator and feedline design. (b)~Side-view of the resonator and the Cu$_2$OSeO$_3$ crystal inside the cavity. Note the different configurations of static external magnetic field $\bm{H}_0$ relative to the oscillating magnetic field $\bm{h}_\mathrm{ac}$.}
		\label{fig:CSO_setup}
	\end{center}
\end{figure}

To verify the expected magnon--photon coupling we used the measurement setup schematically depicted in Fig.~\ref{fig:CSO_setup}(a). We fabricated a loop--gap resonator~\cite{Rinard1999} based on an aluminum cylinder (height $h=\SI{22}{\milli\meter}$, inner diameter $d=\SI{10}{\milli\meter}$, outer diameter $D=\SI{22}{\milli\meter}$). The gap with a width of $g=\SI{0.5}{\milli\meter}$ is filled with a high-dielectric-constant material. A copper loop grounded to the shielding was used as a microwave feedline to inductively couple to the resonator. The feedline was connected to a VNA\ and the complex reflection parameter~$S_{11}$ was measured as a function of the microwave frequency $f$ and the external magnetic field~$\bm{H}_0$. The applied microwave power was \SI{0}{\dBm}, and we confirmed that the magnetization dynamics of the system is in the linear regime. The geometry and placement of the coupling loop was adjusted to maximize the absorbed microwave power by the cavity. The loaded quality factor of the cavity was $\SI{350(26)}{}$ at a resonance frequency $f_\mathrm{res}=\SI{683.8}{\mega\hertz}$.

The Cu$_2$OSeO$_3$ single crystal was mounted on a copper rod and inserted into the cavity (for details see Supplemental Material~\cite{SupInf}, Fig.~S1). The cavity with the sample was placed into a helium-flow cryostat equipped with a 3D vector magnet. The resonator features a clearly defined direction of the oscillating driving field~$\bm{h}_\mathrm{ac}$ with good homogeneity inside the cavity as depicted in Fig.~\ref{fig:CSO_setup}(b) and confirmed by our finite element modeling (see Fig.~S2~\cite{SupInf}). Therefore, it is possible to investigate configurations with $\bm{H}_0 \parallel \bm{h}_\mathrm{ac}$ and $\bm{H}_0 \perp \bm{h}_\mathrm{ac}$ as indicated in Fig.~\ref{fig:CSO_setup}(b). We will refer to the situation $\bm{H}_0 \parallel [110] \parallel \bm{h}_\mathrm{ac}$ as parallel excitation~(PE) and to $\bm{H}_0 \parallel [100] \perp \bm{h}_\mathrm{ac}$ as the orthogonal excitation~(OE). 

\begin{figure}
	\begin{center}
		\includegraphics[]{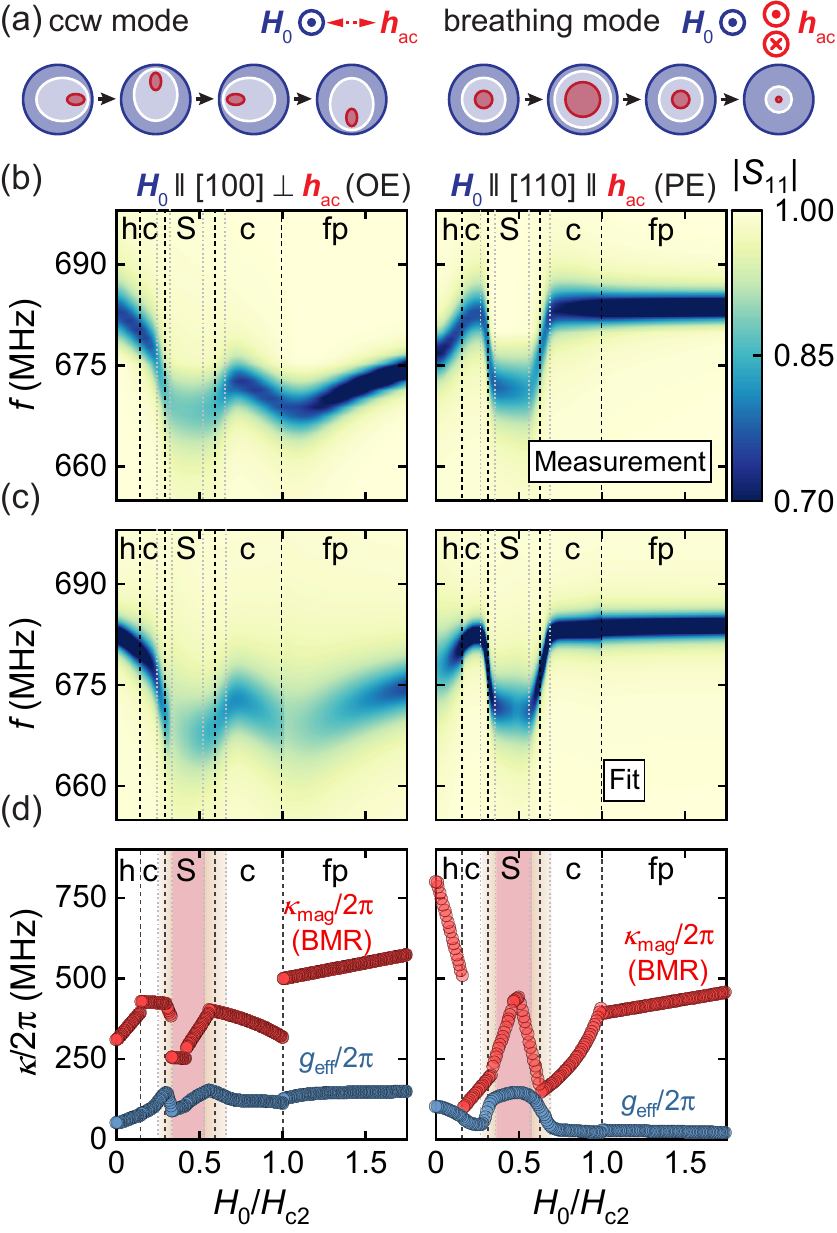}
		\caption{Magnon--photon coupling in a skyrmion host material. (a)~Illustration of the two dominant skyrmion resonance modes. The oscillating driving field $\bm{h}_\mathrm{ac}$ is either orthogonal (ccw mode) or parallel (breathing mode) to the external magnetic field~$\bm{H}_0$. (b)~Measured reflection parameter $|S_{11}|$ as a function of the magnetic field $H_0$ and the applied microwave frequency~$f$. The vertical dashed lines indicate phase boundaries (c.f. Fig.~S4~\cite{SupInf}). The magnetic fields are normalized to the critical field $H_{\mathrm{c}2}$. (c)~Simulated spectrum calculated with Eq.~\eqref{eq:CSO_InputOutput}. (d)~Fitted coupling rate $g_\mathrm{eff}$ extracted from fitting Eq.~\eqref{eq:CSO_InputOutput} to frequency traces at fixed magnetic field and the magnon loss rate $\kappa_\mathrm{mag}$ from broadband magnetic resonance~(BMR) measurements (c.f. Fig.~S6~\cite{SupInf}) used to calculate the spectrum shown in~(c). The fit error bars are smaller than the symbol size. The rather broad phase transition \textbf{S}$\leftrightarrow$\textbf{c} due to the irregular shape of the single-crystal is indicated by dashed gray lines or by the orange shading.}
		\label{fig:CSO_inputoutput_colormaps}
	\end{center}
\end{figure}

The skyrmion lattice forms a periodic hexagonal lattice perpendicular to the external magnetic field~\cite{Muhlbauer2009}. Its breathing mode is excited in PE\ configuration, while the counterclockwise mode dominates in OE\ configuration~\cite{Onose2012, Garst2017} as depicted in Fig.~\ref{fig:CSO_inputoutput_colormaps}(a). All further dynamical modes are also expected to couple in the OE\ configuration. 

Typical results from the reflection measurements performed with the VNA are shown in Fig.~\ref{fig:CSO_inputoutput_colormaps}(b) in form of the background-corrected (see Fig.~S3~\cite{SupInf}) reflection parameter $|S_{11}|$. In order to account for different demagnetization fields, the magnetic field is normalized to the critical field $H_{\mathrm{c}2}$ namely \SI{71.7}{\milli\tesla} for OE\ and \SI{39.1}{\milli\tesla} for PE. For OE, a shift of the cavity resonance frequency is observed in the field-polarized and conical phase in agreement with Fig.~\ref{fig:CSO_general}(d). In the skyrmion lattice phase, we observe coupling between the ccw\ mode and the cavity photons. There is no distinct transition from the conical to the helical phase because the external magnetic field~$\bm{H}_0$ is applied along the magnetic easy axis of the cubic anisotropy in Cu$_2$OSeO$_3$~\cite{Halder2018} where the system does not rearrange in energetically unfavorable domains upon decreasing field but keeps a small net moment along $\bm{H}_0$~\cite{Bauer2017, Kindervater2019, Milde2020}.

If the static magnetic field is parallel to the oscillating driving field (PE), no coupling is observed in the field-polarized and conical phase. As the magnetic moments are aligned with $\bm{H}_0$, the driving field $\bm{h}_\mathrm{ac}$ cannot exert a torque on the precessing magnetic moments. In the skyrmion lattice phase, we clearly observe coupled dynamics of the cavity and the breathing mode of the Cu$_2$OSeO$_3$ spin system. This coupling is most efficient for the driving field $\bm{h}_\mathrm{ac}$ parallel to the static field $\bm{H}_0$ (see Fig.~\ref{fig:CSO_inputoutput_colormaps}(a)). 
In the helical phase, a finite coupling is observed in contrast to the OE\ case. The domain population in the helical phase crucially depends on the field direction as well as on the temperature and field history. For decreasing magnetic field along the $\langle 110 \rangle$-direction, two domains are equally populated leading to a finite angle between them~\cite{Bauer2017, Kindervater2019, Milde2020}. Consequently, a finite coupling in the helical phase is observed.

We now turn to a quantitative evaluation of the magnon--photon coupling employing the input--output formalism in the following manner consisting of three steps: First, we extract the external loss rate~$\kappa_\mathrm{ext}$, the total loss rate~$\kappa_\mathrm{res}$, and resonance frequency~$f_\mathrm{res}$ of the cavity by fitting Eq.~\eqref{eq:CSO_InputOutput} to a frequency trace of the reflection parameter $|S_{11}|$ at the largest static magnetic field available ($\mu_0H_0=\SI{126}{\milli\tesla}$ for PE). The system is assumed to be unperturbed and therefore $g_\mathrm{eff}=0$. The following values are inferred: $f_\mathrm{res}=\SI{683.8}{\mega\hertz}$, $\kappa_\mathrm{res}/(2\pi)=\SI{1.041(4)}{\mega\hertz}$, and $\kappa_\mathrm{ext}/(2\pi)=\SI{0.848(4)}{\mega\hertz}$. The internal loss rate of the cavity $\kappa_\mathrm{int}$ is smaller than the external loss rate $\kappa_\mathrm{ext}$, indicating that the cavity is overcoupled~\cite{Lambert2020}.

In a second step, we keep these parameters fixed and fit Eq.~\eqref{eq:CSO_InputOutput} to all frequency traces of $|S_{11}|$ for a series of external magnetic fields~$H_0$ along both magnetic field directions. From these fits, the coupling strength $g_\mathrm{eff}$ is extracted. The field-dependent resonance frequency of the different magnon modes $f_\mathrm{mag}$ and the magnon decay rate $\kappa_\mathrm{mag}$ is taken from broadband magnetic resonance spectroscopy data measured on the same crystal (see Figs.~S5 and~S6~\cite{SupInf}). Note that the cavity-based technique probes the whole sample volume while the standard broadband technique is sensitive to spin dynamics within the first few \si{\micro\meter} of the sample above the coplanar waveguide. Due to the irregular shape of the crystal, its demagnetization fields are inhomogeneous, which leads to minor deviations in the resonance frequencies and loss rates of the magnon modes and magnetic fields of the phase transition compared to the broadband magnetic resonance technique.

Using the extracted parameters allows to recalculate $|S_{11}|$ using Eq.~\eqref{eq:CSO_InputOutput}, which is displayed in Fig.~\ref{fig:CSO_inputoutput_colormaps}(c). The good agreement between this fit and the measurement in Fig.~\ref{fig:CSO_inputoutput_colormaps}(b) is found. The parameters $\kappa_\mathrm{mag}$ and $g_\mathrm{eff}$ are shown in Fig.~\ref{fig:CSO_inputoutput_colormaps}(d). The magnon decay rate $\kappa_\mathrm{mag}$ is larger than the coupling strength $g_\mathrm{eff}$ in all magnetic phases which classifies the system to be in the Purcell coupling regime with $\kappa_\mathrm{res}<g_\mathrm{eff}<\kappa_\mathrm{mag}$~\cite{Zhang2014}. For PE, the coupling strength outside the skyrmion lattice and helical phase becomes small. In the skyrmion lattice phase, the breathing mode couples to the cavity and we extract almost the same effective coupling rate~$g_\mathrm{eff}$ as for the counterclockwise rotation mode. Note that the resonator properties may subtly evolve with magnetic field, also leading to minor changes in~$g_\mathrm{eff}$. 

\begin{figure}
	\begin{center}
		\includegraphics[]{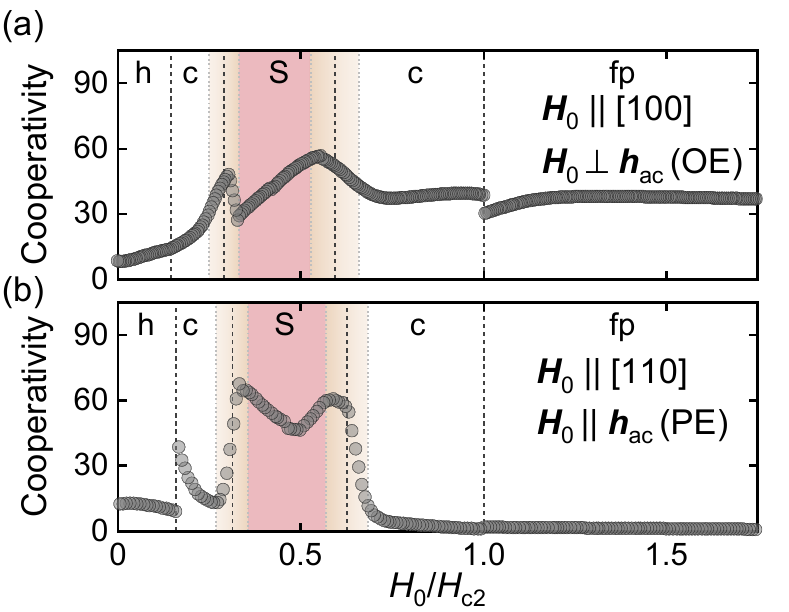}
		\caption{Cooperativity calculated by using Eq.~\eqref{eq:Cooperativity} and the parameters shown in Fig.~\ref{fig:CSO_inputoutput_colormaps}(d). (a)~Orthogonal excitation (OE). (b)~Parallel excitation (PE), for which the cooperativity can be tuned from 1 to 60. The fit error bars are smaller than the symbol size.}
		\label{fig:CSO_cooperativity}
	\end{center}
\end{figure}

In the third and final step, the cooperativity $C$ of the system is calculated as a measure for the coherent exchange of excitations. The cooperativity defined as~\cite{Herskind2009,Huebl2013,Zhang2014,Maier-Flaig2017}
\begin{equation}
C = \frac{g_\mathrm{eff}^2}{\kappa_\mathrm{res}\cdot \kappa_\mathrm{mag}}
\label{eq:Cooperativity}
\end{equation}
is shown in Fig.~\ref{fig:CSO_cooperativity}. For OE, the cooperativity increases from $C\approx 8$ in the helical phase to a high cooperativity of $C\approx 50$ in the skyrmion lattice phase. Similar values for the cooperativity are found in the skyrmion lattice, conical and field-polarized phases away from the phase-transition regions. 

For PE, the cooperativity~$C$ is small or close to unity in the field-polarized and conical phase. A drastic change of the cooperativity is observed when a magnetic phase transition into the skyrmion lattice phase is induced by a small variation of the magnetic field. The cooperativity reaches a maximum value of $C\approx60$ and can be tuned between its maximum and minimum value by changing the external magnetic field by $\sim\SI{10}{\milli\tesla}$. The change of the magnetic field induces a magnetic phase transition which in turn allows us to control the effective coupling rate~$g_\mathrm{eff}$ between the microwave photons and the magnons in Cu$_2$OSeO$_3$. Here, we utilize that the breathing mode exhibits a different excitation geometry compared to the magnon excitations in the other magnetic phases. This property is unique to the topologically protected skyrmion lattice phase. At the phase boundaries, the magnon--photon cooperativity distinctly changes with magnetic field due to the induced phase transition. 
Similarly, at the boundary of the helical phase, when the helices reorientate under decreasing magnetic field and gain components orthogonal to $\bm{h}_\mathrm{ac}$, the microwave photons can more efficiently couple to the spin system. However, such transitions lack the pronounced excitation selectivity.

In conclusion, we experimentally demonstrated a selective coupling between magnetic excitations in the chiral magnet Cu$_2$OSeO$_3$ and photons inside a three-dimensional microwave cavity by using magnetic resonance spectroscopy. The coupling between the topologically protected skyrmions and the photons in the cavity is mediated by the dipolar interaction 
and a high magnon--photon cooperativity is observed. 
By changing the magnetic field in a small field range, the system undergoes a phase transition resulting in a strong tunability of the magnon--photon coupling rate and therefore the magnon-photon cooperativity. The phase change is induced by means of a control parameter, which can either be magnetic field, temperature, or electric field~\cite{SekiPRB2012,White2014,Okamura2016}. Tuning the cooperativity in a dispersive read-out scheme can have great potential in skyrmion racetrack applications~\cite{Fert2013} or in hybrid systems such as magnon--qubit systems~\cite{Lachance-Quirion2019, Wolski2020}.

{\it Acknowledgments.} -- We thank A.~Habel for technical support. This work has been funded by the Deutsche Forschungsgemeinschaft (DFG, German Research Foundation) via the excellence cluster MCQST under Germany's Excellence Strategy EXC-2111 (Project No.\ 390814868). L.L. and M.W. gratefully acknowledge the financial support of the DFG via the projects WE~5386/4-1 and WE~5386/5-1. F.X.H., A.B. and C.P. acknowledge financial support from the DFG under TRR80 (From Electronic Correlations to Functionality, Project No.\ 107745057, Projects E1 and F7) and SPP2137 (Skyrmionics, Project No.\ 403191981, Grant PF393/19) and from the European Research Council (ERC) through Advanced Grants No.\ 291079 (TOPFIT) and No.\ 788031 (ExQuiSid).


%

\end{document}